# Quantum emitter formation in carbon-doped monolayer hexagonal boron nitride

Hongwei Liu[1#], Noah Mendelson[2,3#], Irfan H. Abidi[1,4], Shaobo Li[5], Zhenjing Liu[1], Yuting Cai[1], Kenan Zhang[1], Jiawen You[1], Mohsen Tamtaji[1], Hoilun Wong[1], Yao Ding[6], Guojie Chen[7*], Igor Aharonovich[2,8], Zhengtang Luo[1*]

[1]Department of Chemical and Biological Engineering, Guangdong-Hong Kong-Macao Joint Laboratory for Intelligent Micro-Nano Optoelectronic Technology, William Mong Institute of Nano Science and Technology, and Hong Kong Branch of Chinese National Engineering Research Center for Tissue Restoration and Reconstruction, the Hong Kong University of Science and Technology, Clear Water Bay, Kowloon, Hong Kong, 999077, P.R. China

[2]School of Mathematical and Physical Sciences, University of Technology Sydney, Ultimo, New South Wales 2007, Australia

[3]Pritzker School of Molecular Engineering, University of Chicago, Chicago, IL 60637, USA

[4]Centre for Advanced 2D Materials, National University of Singapore, 117542, Singapore

[5]School of Materials Science and Engineering, South China University of Technology, Guangzhou 510640, P.R. China

[6]School of Materials Science and Engineering, Wuhan University of Technology, Wuhan 430070, China

[7]Guangdong-Hong Kong-Macao Joint Laboratory for Intelligent Micro-Nano Optoelectronic Technology, Foshan University, 528225, China

[8]ARC Centre of Excellence for Transformative Meta-Optical Systems (TMOS), University of Technology Sydney, Ultimo, New South Wales 2007, Australia

*Corresponding Author, email: Guojie Chen: gjchen@fosu.edu.cn; Zhengtang Luo: keztluo@ust.hk.

#These authors contributed equally.

**Abstract**

Single photon emitters (SPEs) in hexagonal boron nitride (hBN) are promising candidates for quantum light generation. Despite this, techniques to control the formation of hBN SPEs down to the monolayer limit are yet to be demonstrated. Recent experimental and theoretical investigations have suggested that the visible wavelength single photon emitters in hBN originate from carbon-related defects. Here we demonstrate a simple strategy for controlling SPE creation during the chemical vapor deposition growth of monolayer hBN via regulating surface carbon concentration. By increasing surface carbon concentration during hBN growth, we observe increases in carbon doping levels by 2.4 fold for B-C bonds and 1.6 fold for N-C bonds. For the same samples we observe an increase in SPE density from 0.13 to 0.30 emitters/$\mu m^2$. Our simple method enables the reliable creation of hBN SPEs in monolayer samples for the first time, opening the door to advanced 2D quantum state engineering.

**Key words**: hBN, 2D Materials, carbon doping, chemical vapor deposition, monolayer, single photon emitters

**Introduction**

Solid-state single photon emitters (SPEs) are fundamental building blocks for many emerging quantum technologies.[1-3] To date, solid-state SPEs have been discovered in materials of different dimensionalities such as zero-dimensional materials (GaAs, InGaAs quantum dots), one-dimensional materials (carbon nanotube, InP nanowires), and three-dimensional wide band gap materials (diamond, GaN).[4] Recently, SPEs in two dimensional (2D) materials have been pushed to the forefront,[6] displaying extraordinary advantages including high photon extraction efficiency,[5] and they offer straight-forward integration into on-chip architectures.[6, 7] Within this subclass, defects in the wide band-gap hexagonal boron nitride (hBN) are especially exciting as they display outstanding optical and spin properties at room temperature.[8-15] hBN SPEs are highly tunable,[16-18] display high optically detected magnetic resonance contrast,[11] and have been successfully integrated with optical cavities and waveguides through both hybrid and monolithic approaches.[19]

Interestingly, hBN SPEs have been studied almost exclusively in bulk hBN samples, from a few layers up to hundreds of nanometers thick. In fact, study of SPEs embedded in monolayer (ML) hBN has remained limited to probing their blinking dynamics through wide-area imaging techniques, as reliable fabrication techniques for SPE generation at the monolayer (ML) limit have yet to be demonstrated. Efficient SPE creation in ML hBN presents a critical hurdle for the platform, as the potential for 2D quantum state engineering—such as integration with van der Waals heterostructures[20] or through controlling the twist angle between adjacent layers[21]— is just beginning to be explored.

A longstanding bottleneck for the fabrication of ML hBN SPEs has been a lack of knowledge about the chemical nature of the emissive defect. Recently, direct experimental evidence was reported suggesting carbon heteroatoms are responsible for SPEs in the visible region.[22-24] Furthermore, control over the SPE density from isolated singles to dense ensembles was demonstrated via controlled carbon doping during bottom-up growth of ~40 nm hBN films.[25] These results highlighted the potential for employing similar techniques to reliably generate ML hBN SPEs during bottom-up growth. One appealing bottom-up technique frequently utilized for 2D materials growth is chemical vapor deposition (CVD). The CVD method possess intrinsic advantages for targeted applications, such as synthesis of large-area films with controlled thickness down to monolayer.[25-29] Additionally, controlling CVD growth conditions has proved effective in modulating the emission properties of incorporated SPEs,[30] and enabled the direct growth of hBN SPEs on prefabricated dielectric substrates such as pillars and waveguides.[12, 31]

In this work, we demonstrate a CVD based method for controlling the level of carbon doping in ML hBN crystals by introducing methane during growth, and for the first time reliably generating SPEs in hBN MLs. We additionally demonstrate a technique to reduce intrinsic carbon concentration below standard levels during CVD growth using *molten Cu* as the growth substrate. The structural differences and C doping levels are characterized by atomic force microscopy (AFM), Raman spectroscopy, X-ray photoelectron spectroscopy (XPS), and Time-of-flight second ion mass spectroscopy (ToF-SIMS). Confocal and wide-field photoluminescence experiments are used to correlate the material properties to SPE formation. We find SPEs are present in each as-grown type of hBN ML crystals, and record $g^{(2)}<0.5$ in each case. We establish a clear correlation between the amount of carbon doping of hBN ML crystals and the density of corresponding SPEs incorporated. We propose a simple model based on the concentration of surface carbon during growth to explain our results. Our work provides an effective approach to control SPEs formation during CVD hBN growth and will enable further study of the optical properties of hBN SPEs in MLs.

**Experimental sessions**

*Chemical vapor deposition growth of three kinds of hBN.* 2 mg ammonia borane (Aladdin, 97%) was used as the growth precursor and is contained in a quartz boat. For the growth of $C_{0.20}$-hBN, a piece of Cu foil (Alfa Aesar, #13382) was chemically polished in Acetic acid for 5 min, following with rinsing in deionized (DI) water and drying by Nitrogen gun. The Cu foil was placed on a quartz plate and loaded into a 1-inch tube furnace. Ammonia borane was loaded at the upstream and heated by a heating coil. After purging the growth system with 200 sccm Argon gas, the furnace was heated up to 1045 °C in 30 minutes. Then, the Cu foil was annealed with mixed gas of 200 sccm Argon and 20 sccm Hydrogen for 30 minutes. The hBN growth started with heating up the ammonia borane to 95 °C and grew for a certain duration. Finally, the growth was finished by powering off the furnace and heating coil, and turning off the hydrogen gas. The system was cooling down to room temperature with flowing Argon. The growth of $C_{0.30}$-hBN was similar to the growth of $C_{0.20}$-hBN, while 20 sccm methane was simultaneously introduced when sublimating ammonia borane. For $C_{0.13}$-hBN growth, three pieces of 1.8 cm×1.8 cm Cu foil were stacked together with a tungsten foil support and the growth temperature was increased to 1085 °C. After growth, the as-grown samples were heated up to 200 °C for 10 s to visualize the grains.

*Transfer of as-grown hBN.* hBN crystals were transferred by a PMMA-assisted bubble transfer method. The target hBN on Cu foil was first cut in to 1 cm×1 cm size and spin-coated with PMMA (1000 rpm, 10 s and then 3000 rpm, 60 s) and then worked as cathode, while a Pt foil was used as anode and they are connected to a DC power supply. 1M NaOH solution was prepared as the electrolyte. The hBN/PMMA was gradually separated from the Cu substrate by the generating $H_2$ bubble at 2.3 V. The the hBN PMMA was transferred and floated onto DI water to remove the adsorbed sodium ions. Then the hBN/PMMA was scooped up with Si wafer and dried on a hot plate at 75 °C. Finally, the PMMA was removed by sinking into boiling acetone.

*Structural characterizations of hBN.* Optical microscope (LEICA DMLM optical microscope) was used to directly observe the as grown hBN samples. Macroscopical structure was revealed by Raman spectroscopy (Renishaw Raman RM3000 scope) with a 514 nm source. The diffusions of elements into Cu substrates were investigated by Time-of-Flight Secondary Ion Mass Spectrometry (ToF-SIMs, ION-TOF GmbH, Münster, Germany) using $Cs^+$ ion beam for sputtering. The thickness of hBN films were measured by Atomic Force Microscopy (AFM, Bruker Innova, Laser Diode 670 nm, Class b). X-ray photoelectron spectroscopy (XPS, Thermal Fisher, ESCALAB 250Xi) was conducted for element state analysis.

*Photoluminescence measurements.* Confocal PL was performed using a custom-built scanning confocal microscope, and exciting via a continuous wave (CW) 532 nm laser (Gem 532, Laser Quantum Ltd.). The laser was directed to a high numerical aperture (100×, NA = 0.9, Nikon) objective lens, and scanning was done with an X−Y piezo fast steering mirror (FSM-300). The collection path was filtered with a 532 nm dichroic mirror (532 nm laser BrightLine, Semrock) followed by a long pass 568 nm filter (Semrock), before coupling into a graded-index multimode fiber (0.22 NA) with a fiber aperture of 62.5 μm acting as a confocal pinhole. A flip mirror directed the emission to a spectrometer (Acton Spectra Pro, Princeton Instrument Inc.) or to two avalanche photodiodes (Excelitas Technologies) in a Hanbury-Brown and Twiss configuration. Correlation measurements were performed with a time-correlated single photon counting module (PicoHarp 300, PicoQuant).

*Wide-field-imaging*. Wide-field imaging experiments utilized a lens system to focus the excitation beam at the Fourier plane of the objective, creating a collimated illumination spot enabling the collective excitation of a large area of the sample. The emission from various hBN samples was recorded on an EMCCD (Andor iXon Ultra 888) with an exposure time of 200 ms.

**Results and discussions**

hBN monolayer crystallites are grown *via* atmospheric pressure chemical vapor deposition (APCVD) with ammonia borane as the growth precursor and Cu foils as the growth substrate (The experimental setup is shown in **Figure S1**. See experimental section for details). To illustrate the role of carbon atoms on the formation of SPEs, we synthesize three samples with different C doping levels. The first sample type is hBN grown on *molten Cu* to absorb the intrinsic carbon impurities, labeled as "$C_{0.13}$-hBN". The subscript "0.13" is related to the density of generated single photon emitters, which will be discussed later. Even though there is no artificially carbon introduction, there remains unavoidable intrinsic carbon impurities available during growth, especially in the industrial Cu foil where C can diffuse from the bulk to the surface during growth, a process known to influence graphene nucleation.[32, 33] Therefore, we utilize *molten Cu* on a W support as a growth substrate since *molten Cu* has a higher carbon solubility than solid Cu.[34, 35] As a result, molten Cu can absorb residual carbons and avoid C segregation onto the Cu surface. W plate is utilized as the *molten Cu* support rather than quartz plate to avoid the shrinking of Cu liquid into Cu spheres.[36] The furnace temperature is set slightly higher than the melting point of Cu (1085°C) for 5 min before hBN growth to achieve a completely molten Cu surface as the growth substrate. The second sample type is termed "$C_{0.20}$-hBN," taking place on a solid Cu foil with a quartz support at 1045°C. The third sample type has increased C doping, achieved by introducing a 20sccm methane flow during the otherwise procedurally identical hBN growth. This case is labeled as "$C_{0.30}$-hBN". Growth is performed for 5 min in each case.

We first investigate the growth behaviors of these three cases. To enable easy optical visualization of as-grown hBN domains, the Cu foils are briefly heated up to 200°C to visualize the hBN crystals, **Figure 1a~c**. The morphologies of the resulting hBN crystals vary significantly across the three growth protocols. $C_{0.13}$-hBN grains grown on *molten Cu* tend to be circular in shape and uniformly distributed on the *molten Cu* surface. The circular shape results from the kinetics dominant-growth regime of hBN on *molten Cu,* where the B and N atoms can attach to the less reactive sites of hBN crystals to form circular hBN crystals. In contrast, we observe triangular shaped domains for hBN grown on solid Cu substrates, consistent with previous literature,[37] and attributed to the more energetically stable N-terminated edges of hBN on solid Cu substrate.

In addition, different growth kinetics are also observed in these three cases. The insets of **Figure 1a~c** display the size distribution**.** Here, the size of the hBN grains are defined as their areas. $C_{0.13}$-hBN growth on *molten Cu* shows enhanced growth kinetics and is attributed to the higher growth temperature[38] with grain sizes ranging from 25~35 $\mu m^2$. The size of $C_{0.20}$-hBN grains ranges from 2~2.5 $\mu m^2$, shown in **Figure 1b**. Interestingly, on solid Cu, the introduction of methane increases the growth rate of hBN while decreasing the nucleation density, giving $C_{0.30}$-hBN grains of 5~15 $\mu m^2$ shown in **Figure 1c**. Similar results were observed in reported work for hBN grown on carburized Fe, where carbon was utilized as a tool to reduce the residual oxygen level.[39]

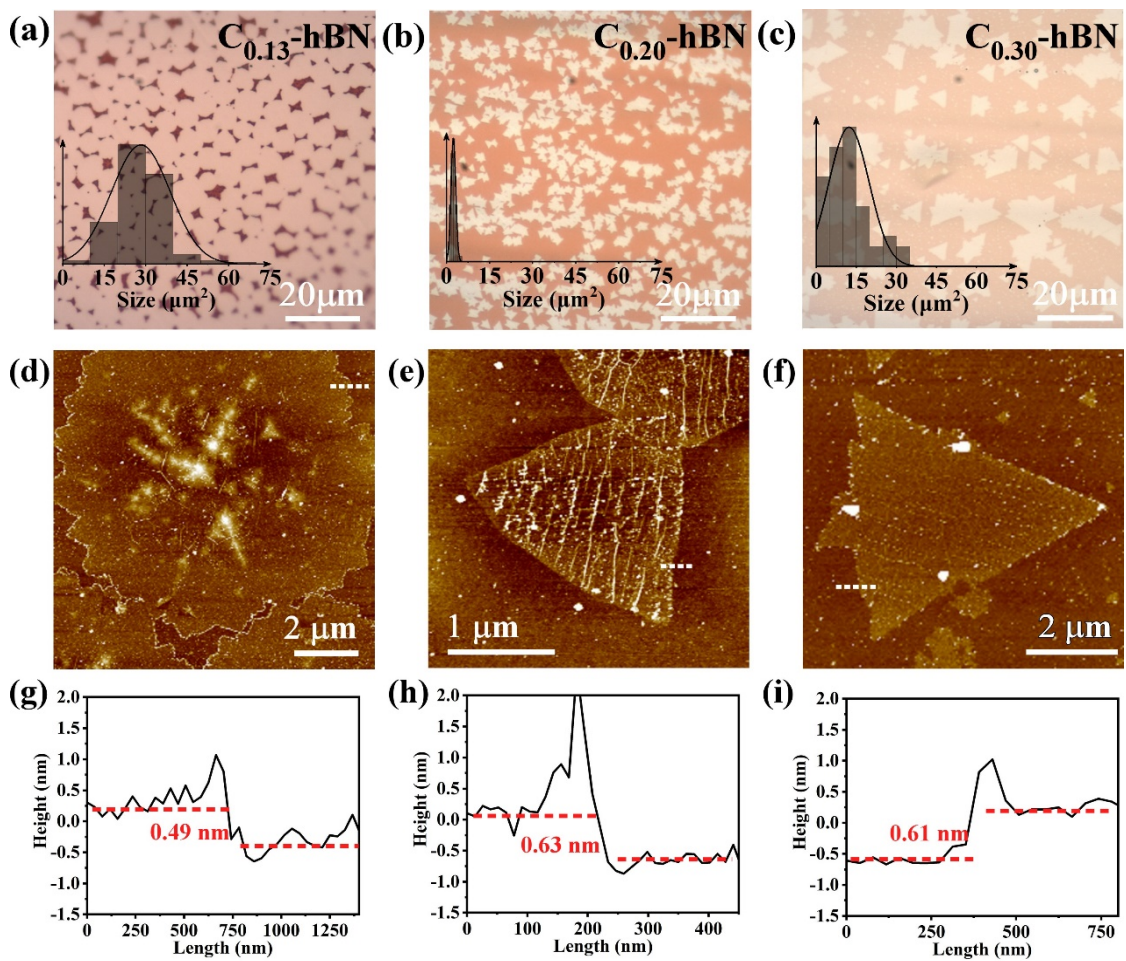

**Figure 1. Morphologies of hBN grains.** Optical images of as-grown hBN with (a) $C_{0.13}$-hBN, (b) $C_{0.20}$-hBN and (c) $C_{0.30}$-hBN after visualization with the size distributions as insets. AFM measurements and corresponding line profiles of (d, g) $C_{0.13}$-hBN, (e, h) $C_{0.20}$-hBN and (f, i) $C_{0.30}$-hBN crystals. The hBN crystals are monolayers in each case, while there are adlayer nucleation sites on $C_{0.30}$-hBN grown on *molten Cu*.

We then analyze the thickness and surface morphology of each hBN type by atomic force microscopy (AFM). The as-grown samples were transferred from Cu onto Si wafers capped with 280 nm $SiO_2$ (**Figure S2**, See Experimental Sections for details). **Figure 1d~f** displays the AFM map for the different hBN samples, and corresponding line profiles are shown in **Figure 1g~i**. All three methods yield monolayer hBN crystallites, with thicknesses of 0.49 nm, 0.61 nm, and 0.63 nm for $C_{0.13}$-hBN, $C_{0.20}$-hBN, and $C_{0.30}$-hBN, respectively.[40] Notably, we observe that there are always adlayer nucleation sites for $C_{0.13}$-hBN grains grown on *molten Cu*. In our recent work, a kinetic

2D diffusion-reaction model was established for hBN grown on *melt-Cu,* revealing the formation mechanism of these centered-aggregation features of adlayers, where random nucleation and random diffusion are the key factors.

Next we investigate the structural differences by Raman Spectroscopy. **Figure 2a and 2b** display the obtained spectra fit with a single Lorentzian. The peaks of hBN grown by all three methods all locate near 1370 cm$^{-1}$ corresponding to $E_{2g}$ vibration mode of monolayer hBN. The peak positions of $C_{0.13}$-hBN and $C_{0.20}$-hBN are similar at 1370.3 cm$^{-1}$ and 1370.5 cm$^{-1}$, respectively. However, the peak position of $C_{0.30}$-hBN shows a slight red-shift in the spectrum to 1369.6 cm$^{-1}$. We attribute the small red-shift to the incorporation of extra carbon atoms influencing the lattice vibrations of hBN. The full width at half maximum (FWHM) is an important parameter to reveal the crystallinity of materials. The FWHM is 13.8cm$^{-1}$ for $C_{0.13}$-hBN, increasing to 15.5 cm$^{-1}$ (15.8 cm$^{-1}$) for $C_{0.20}$-hBN ($C_{0.30}$-hBN). Our results show that $C_{0.13}$-hBN hBN grown on *molten Cu* displays the highest crystallinity, and decreases as carbon concentration in the chamber rises, as expected for increasingly doped materials.

To confirm the successful incorporation of carbon atoms within the hBN lattice, we analyze the chemical composition of the as-grown hBN monolayer crystals. X-ray photoelectron spectroscopy (XPS) is used to probe the chemical bonding profile in each hBN type by looking at the B1s and N1s core electrons. **Figure 2c and 2d** display percentages of boron-carbon bond (B-C%) and nitrogen-carbon bonding (N-C%). Spectra and peak fits used to evaluate bonding percentages are displayed in **Figure S3**. By analyzing boron and nitrogen bonding, we determine the B-C (N-C) bonding percentage increases from 7% (9%) for $C_{0.13}$-hBN, to 13% (13%) for $C_{0.20}$-hBN and to 17% (14%) for $C_{0.30}$-hBN. We note that the monolayer nature of the grown hBN makes XPS results extremely challenging, and prohibits pre-XPS cleaning methods such as etching some of the material with a gentle Ar beam,[22] and may lead to an overestimate of relative carbon bonding percentages. However, this does not affect the general trend observed, increasing carbon concentration during growth leads to increased carbon doping of the as-grown hBN lattice. These results confirm the role of *molten Cu* as carbon getter to substantially reduce the incorporation of carbon into hBN crystals through a simple experimental modification. Additionally, the XPS results unambiguously confirm that carbon incorporation can be enhanced by introducing methane as carbon source during the reaction, where gas flow provides a simple control over the level of external dopants like carbon.

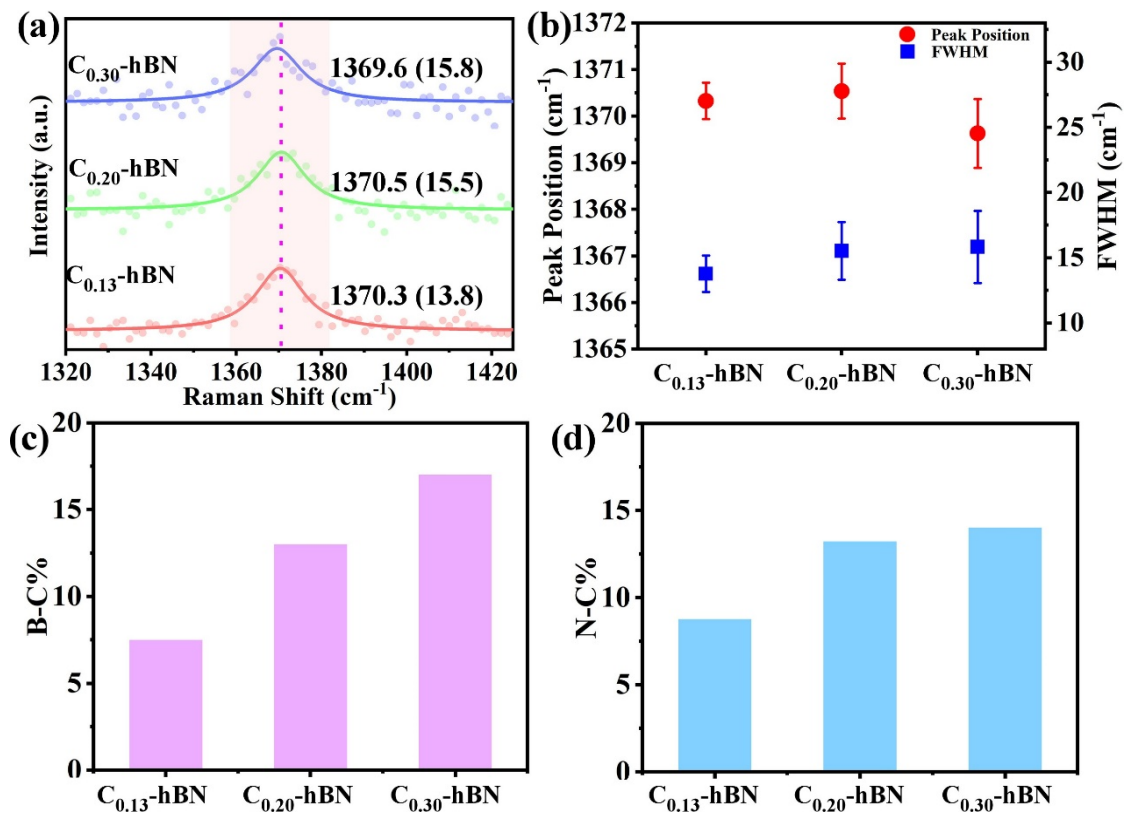

**Figure 2. Structural and composition analysis of as-grown hBN crystals.** (a) Raman spectra of $C_{0.30}$-hBN (blue), $C_{0.20}$-hBN (green) and $C_{0.13}$-hBN (red) with peak positions and FWHMs (in brackets) labeled. A slight blue-shifted is observed with for hBN with higher carbon concentration. (b) Fitting results of Raman spectra in (a), showing that $C_{0.13}$-hBN has the highest crystallinity. The percentages of (c) boron and carbon bond and (d) nitrogen and carbon bond are extracted from the B1s and N1s XPS spectra in **Figure S1** after multipeak fitting.

We next propose a simplified mechanism to understand the modulation of carbon incorporation during hBN growth by our approach. **Figure 3a** displays a schematic representation of our proposed growth model. When solid Cu foil is applied as the growth substrate for hBN growth, the carbon will preferentially aggregate on the metal surface due to low carbon solubility in Cu.[41] This leads to a sizeable C surface concentration level during growth, which can be readily incorporated as C based defects during hBN growth. This C surface concentration only increases when introducing gaseous carbon like methane. On the contrary, when hBN growth is performed on *molten Cu*, the intrinsic carbon impurities are more likely to be dissolved into the bulk Cu catalyst, given the higher solubility of C in Cu at 1085°C,[42] thereby decreasing overall surface carbon levels and thus the observed carbon doping level in the growing hBN crystal.

To further verify our model, we perform Time-of-Flight Secondary Ion Mass Spectrometry (ToF-SIMs) on all three as-grown hBN sample types. **Figure S4** shows the depth profiles of $C^-$ and $CN^-$ species in the results of three cases. For $C_{0.20}$-hBN and $C_{0.30}$-hBN where solid Cu foil is used, carbon atoms diffuse into Cu foil in the depth of only 8~10 nm, indicating that most of the carbon atoms segregate on the Cu surface and can participate in the hBN growth. However, for $C_{0.13}$-hBN, carbon atoms are detected until the depth of 15nm, confirming that *molten Cu* indeed dissolves more carbon impurities and importantly, to a depth where they are less able to participate in hBN growth. To quantitively analyze the results, we consider the diffusion of carbon atoms into Cu substrates as one-dimensional diffusion, where we can apply Fick's second law to fit the ToF-SIMS results to obtain diffusion coefficient $D_C$[43]:

$$\frac{\partial C}{\partial t} = D \frac{\partial^2 C}{\partial x^2} \qquad (1)$$

where $C$ is the concentration of carbon diffused into Cu substrate at depth x and after time t. To solve the equation, we set below boundary conditions:

a. $C_x = C_s$ for $x = 0$, where $C_s$ is the surface carbon concentration

b. $C_x = C_0$ for $x = \infty$, where $C_0$ is the carbon concentration in the bulk Cu

Based on the ToF-SIMS results in **Figure S4**, the carbon concentration in the bulk of Cu substrate is extremely low. Therefore, we can assume $C_0 = 0$. With conditions *a* and *b*, *eq 1* can be integrated into the following equation[44]:

$$\frac{C_x}{C_s} = 1 - \text{erf}\left(\frac{x}{\sqrt{4Dt}}\right) \qquad (2)$$

where $f(x) = \text{erf}(x)$ is the Gaussian error function:

$$f(x) = \text{erf}(x) = \frac{2}{\sqrt{\pi}} \int_0^x e^{-u^2} du$$

Considering *eq 2*, we replotted the ToF-SIMS data in **Figure S4** as $C_x/C_s$ versus

carbon diffusion depth $x$ as shown in **Figure 3b and 3c**. After fitting the curves and taking diffusion time $t$ as the reaction time, i.e., 300s, the carbon diffusion coefficients obtained for different growth methods and from different detected ions can be obtained and the results are summarized in **Table 1**. The obtained $D_C$ of carbon for $C_{0.13}$-hBN was estimated to be ~$7.7 \times 10^{-16} cm^2/s$ measured from C$^-$ ions spectrum and ~$1.1 \times 10^{-15} cm^2/s$, about twice compared with that of carbon for $C_{0.20}$-hBN and $C_{0.30}$-hBN. These results indicate that increased catalyst solubility for C lowers the carbon concentration on the surface, reducing overall C doping levels in the growing hBN. Interestingly, the data also suggest that the methane introduced during $C_{0.30}$-hBN growth compared to $C_{0.20}$-hBN growth does not significantly increase the carbon concentration within the catalyst found in our ToF-SIMS measurement. We note that the ToF-SIMS measurement requires the system to first be cooled down, which can allow atoms to segregate to surface as the temperature (and solubility) of the Cu decreases. However, the lack of any increase in the carbon concentration suggests the carbon introduced during $C_{0.30}$-hBN growth stays either on the Cu surface or the near surface, increasing the probability of being incorporated as a defect during hBN growth.

**Table 1. Carbon diffusion coefficients into Cu substrates from Tof-SIMS (cm$^2$/s).**

| Measured ions | $C_{0.13}$-hBN | $C_{0.20}$-hBN | $C_{0.30}$-hBN |
|---|---|---|---|
| C$^-$ | 7.7*10$^{-16}$ | 3.6*10$^{-16}$ | 4.2*10$^{-16}$ |
| CN$^-$ | 1.1*10$^{-15}$ | 3.9*10$^{-16}$ | 4.3*10$^{-16}$ |

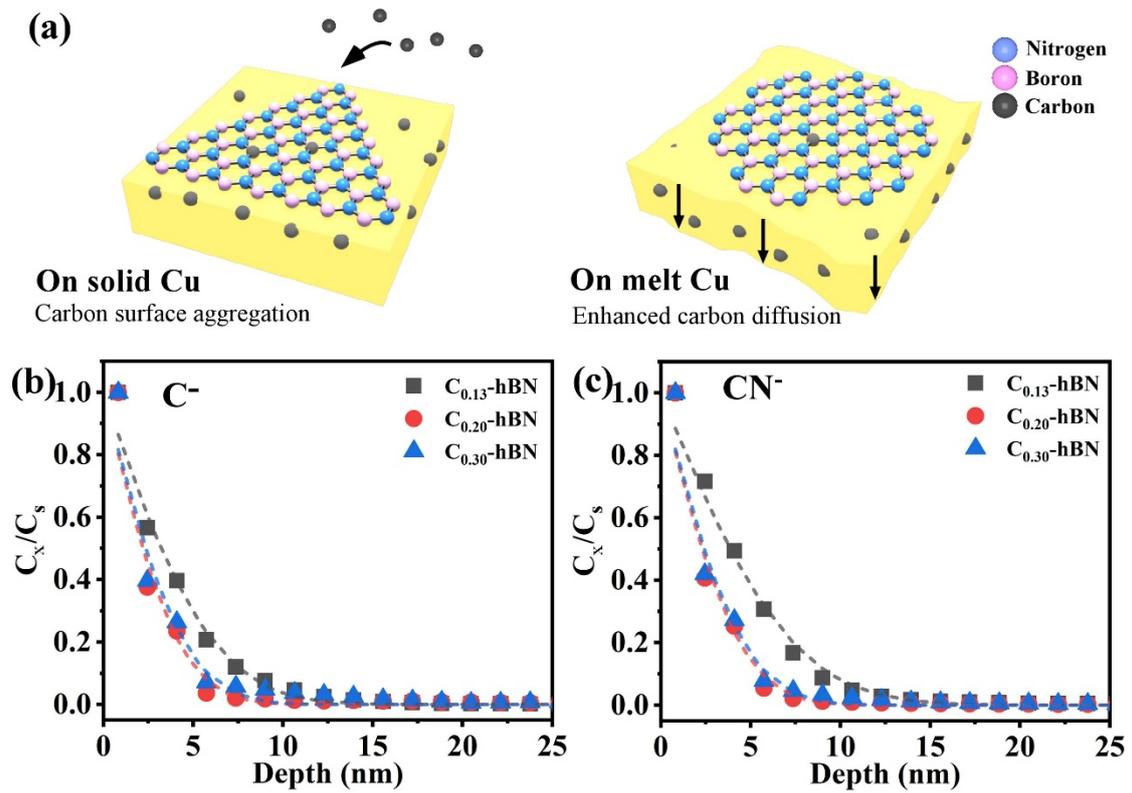

**Figure 3. Controlling carbon surface concentrations during growth.** (a) Schematic model of hBN growth with methane introduction on solid Cu (Left) and applying molten Cu as the carbon impurities absorbent (right). Replotted $C_x/C_S$ versus $x$ from element depth analysis of carbon species (b) C$^-$ ions and (c) CN$^-$ ions by Tof-SIMs. The dash lines are fitted curve. The results confirm the role of *molten Cu* as the absorbent of carbon impurities.

Having established a clear protocol to modulate carbon doping levels in monolayer hBN during CVD growth, we next analyze the optical properties of these three samples to establish a potential link between increased carbon dopants and SPE density. We perform confocal photoluminescence (PL) analysis of each as-grown hBN sample type. PL is performed on a custom-built setup with hBN samples on Si/SiO$_2$ substrates, and excited with a 532 nm continuous wave (CW) laser (see methods). **Figure 4a~c** display the confocal maps of the three sample types, where individual hBN monolayer crystallites can be seen in each case. As seen in the optical images, we find the $C_{0.13}$-hBN and $C_{0.30}$-hBN show larger hBN crystallites in general, while the $C_{0.30}$-hBN displays a noticeable increase in brightness and localized bright spots within in the sample.

**Figure 4d~f** display representative spectra from localized SPEs in each sample

type, as well as corresponding 2$^{nd}$ order autocorrelation measurements confirming the quantum nature of the emission measured in each case ($g^{(2)}(0)<0.5$). For all three sample types we observe zero-phonon lines, and accompanying phonon sidebands separated by ~160 meV, consistent with previous reports.[45] SPEs display emission energies between 600-650 nm consistently across the samples. Emission lines in the samples are quite unstable due to the monolayer nature of the host crystal, however, this could be mitigated in future work through incorporation into van-der Waals heterostructures or other thin-layer capping.

To compare the relative density of SPEs between the three growths we perform wide-field microscopy, a powerful technique allowing for the direct observation of isolated SPEs simultaneously over large areas. **Figure 4g~i** display the wide-field images of the three respective hBN sample types. We observe SPE density increases with the carbon doping level in the respective hBN crystallites. Specifically, we find SPE densities of ~0.13 SPE/µm$^2$, ~0.20 SPE/µm$^2$, and ~0.30 SPE/µm$^2$ for the $C_{0.13}$-hBN, $C_{0.20}$-hBN and $C_{0.30}$-hBN, respectively. Our results provide further evidence for the role of carbon in the creation of visible region SPEs in hBN during CVD growth, and yield the first reliable protocol for creating hBN SPEs in monolayers.

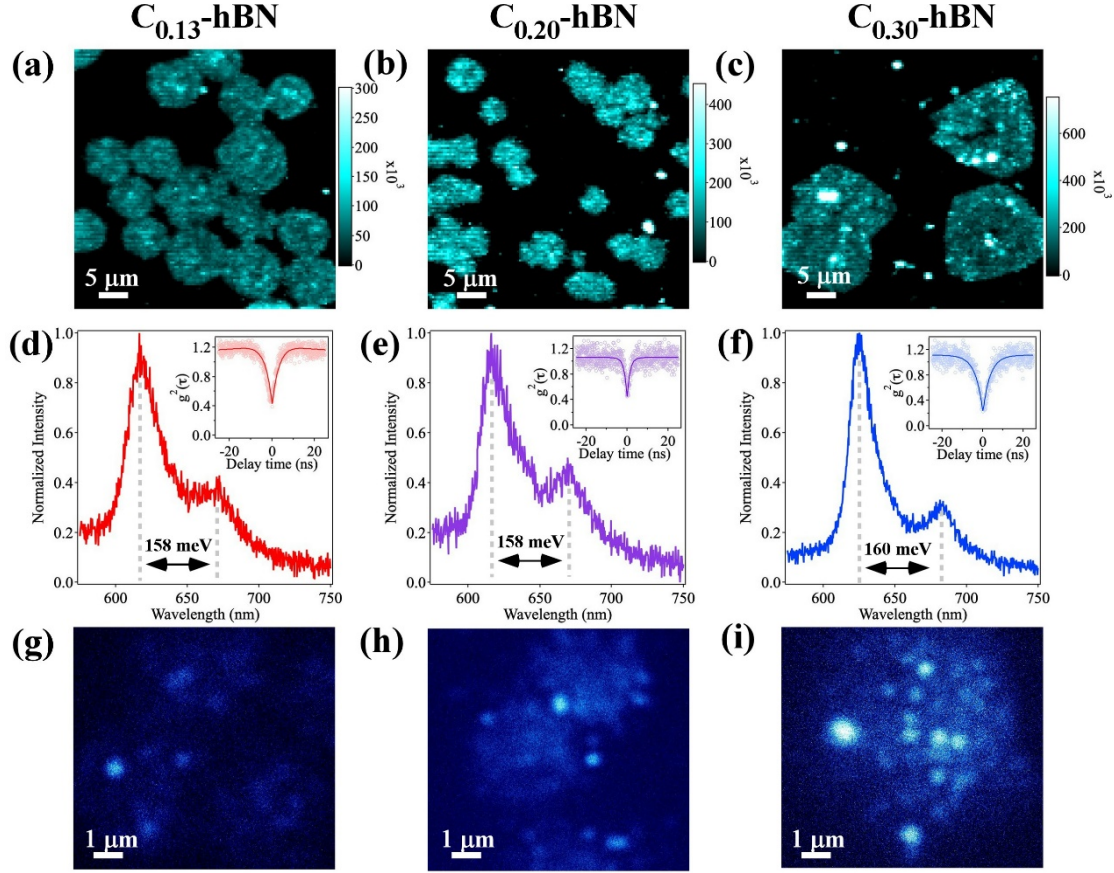

**Figure 4. PL analysis of hBN monolayers.** a, d, and g are recorded from $C_{0.13}$-hBN. b, e, and h were recorded from $C_{0.20}$-hBN. c, f, and i are recorded from $C_{0.30}$-hBN. All data is recorded after transfer to $SiO_2$. (a-c) Confocal scans of each hBN sample type. (d-f) Representative spectra for localized SPEs recorded in each sample type. Inset: $g^{(2)}(t)$ recorded from each SPE. (g-i) Wide-field images of each hBN sample type.

**Conclusion**

In summary, we demonstrated a simple yet effective approach to modulate carbon doping during CVD hBN growth, enabling the efficient creation of SPEs in hBN monolayers. The carbon doping level can be increased by introducing methane during hBN growth or decreased by applying the *molten Cu* approach which reduces observed carbon doping levels. XPS results suggest a ~2.4 (~1.6) fold increase in B-C (N-C) bonding between the $C_{0.13}$-hBN and $C_{0.30}$-hBN. We propose a simple mechanistic understanding of these results, identifying surface carbon concentration levels during growth as the key parameter for C-doping of the growing hBN crystallites. ToF-SIMS analysis confirmed the *molten Cu* approach leads to higher carbon concentrations deeper within the catalyst, supporting the proposed mechanism. Our results offer a simple approach to control heteroatom doping during CVD growth and could be extended to dopants beyond carbon.

Monolayer crystallites grown by all three methods display isolated SPEs ($g^{(2)}(0)<0.5$) with ZPLs ranging from 600-650 nm. SPE density was found to increase with C-doping from ~0.13 to 0.30 SPE/μm$^2$. Our results display a clear correlation between carbon doping and SPE density during CVD growth, corroborating recent works linking carbon incorporation to SPE creation during alternative hBN growth methods.[23, 25] The ability to control hBN thickness and C doping simultaneously enabled the efficient creation of hBN SPEs at the monolayer limit for the first time. Our results will accelerate the characterizations of SPEs in monolayer hBN and enable their integration with van der Waals heterostructures to explore exotic phenomena unique to 2D quantum systems.

**Conflicts of interest**

These authors respectfully declare that, there are no conflicts of interest to acknowledge for this research.

**Acknowledgement**

Z.L. acknowledge supports by the NSFC-RGC Joint Research Scheme (N_HKUST607/17), and the IER foundation (HT-JD-CXY-201907), "International science and technology cooperation projects" of Science and Technological Bureau of Guangzhou Huangpu District (2019GH06), Guangdong Science and Technology Department (Project#:2020A0505090003), Shenzhen Special Fund for Central Funds Guiding the Local Science and Technology Development (2021Szvup136) and Research Fund of Guangdong-Hong Kong-Macao Joint Laboratory for Intelligent Micro-Nano Optoelectronic Technology (No. 2020B1212030010). Technical assistance from the Materials Characterization and Preparation Facilities of HKUST is greatly appreciated.